# Urban Energy Flux: Human Mobility as a Predictor for Spatial Changes

Neda Mohammadi[1], and John E. Taylor[2]*

As a key energy challenge, we urgently require a better understanding of how growing urban populations interact with municipal energy systems and the resulting impact on energy demand across city neighborhoods, which are dense hubs of both consumer population and $CO_2$ emissions. Currently, the physical characteristics of urban infrastructure are the main determinants in predictive modeling of the demand side of energy in our rapidly growing urban areas; overlooking influence related to fluctuating human activities. Here, we show how applying intra-urban human mobility as an indicator for interactions of the population with local energy systems can be translated into spatial imprints to predict the spatial distribution of energy use in urban settings. Our findings establish human mobility as an important element in explaining the spatial structure underlying urban energy flux and demonstrate the utility of a human mobility driven approach for predicting future urban energy demand with implications for $CO_2$ emission strategies.

The earth's rapidly expanding urban spaces are growing in terms of both technology and population at a rapid rate, creating the most complex built environments in human history. A 2014 United Nations report announced that 54% of the world's population now resides in urban areas[1]. It was not until 1950 that New York became the world's first megacity with a population of 10 million or more inhabitants[2], but over the following decades others joined the category and today's 28 megacities are projected to increase to 41 by 2030[1]. A growth of this magnitude has significant implications for global energy, as urban areas are major consumers (up to 80%) of the world's total energy production, and an increase of up to 56% in global energy requirements has been predicted between 2010 and 2040[3]. Managing and allocating resources and generating credible predictions of future energy demand requires a clear understanding of the spatial distribution and patterns of urban energy consumption by identifying the factors and indicators that determine and influence the demand side of energy.

The spatial distribution of energy use in urban areas depends on human activities and people's daily routines. Certain types of energy use behavior are clustered in specific spatial and temporal locations[4]. These include work, home and leisure activities, all of which have an impact on future energy demand in distinct areas of the city. For example, individuals may practice low consumption habits at work but then consume disproportionate amounts of energy later in the day when they arrive home and they may be consuming energy from either exclusive or shared resources. It is thus important to identify the drivers of this consumption in different regions and explore the patterns and predictors of urban energy use. Unreliable predictions and poor management decisions about future patterns of energy consumption and demand due to non-quantified human dimensions of energy use may adversely affect cities' energy resilience, leading to enormous waste in the financial resources municipalities invest in energy distribution and infrastructure.

## Urban Energy Spatial Flux and Demand Prediction

In predicting future energy demand, spatial patterns currently tend to be primarily characterized in terms of physical determinants of the urban infrastructure such building types[5,6], city location and district features[7]; and building age and function[8], generally also taking into account external conditions such as weather and geographic location[9]. However, given that the planet's urban population is predicted to rise by an additional 2.5 billion inhabitants in urban environments by 2050[1], the scale and diversity of the human activities driving energy consumption continues to expand[10]. The spatial distribution of energy use thus remains in a continuous state of flux and the resulting intricate interdependencies between infrastructure, services, and individuals presage an ambiguous future in which we will face challenges of which we are not yet aware. This means that existing approaches, which are principally based on the physical characteristics of urban infrastructure, will fail to reliably explain patterns of urban energy, and lead to widely inaccurate predictions of energy demand. This raises important questions regarding our ability to create and maintain adequate energy resources to meet demand in our large and growing population centers.

Although several studies have recognized that different human activity patterns may be responsible for fluctuations in energy consumption[11,12], researchers have only captured this effect within limited areas, such as individual buildings, which cannot adequately represent the global patterns and structures of energy consumption at an urban level. Much of our current understanding of future patterns of energy use comes from decades of research focusing specifically on the physical properties of cities, omitting any consideration of quantified measures of human activities. Despite the importance of the role urban populations play in the transformation of energy systems[13], reflections of their fluctuating activities are largely absent from urban energy studies.

[1] The Charles E. Via, Jr. Department of Civil and Environmental Engineering, Virginia Tech, Blacksburg, VA 24060.  [2] Frederick Law Olmsted Professor, School of Civil and Environmental Engineering, Georgia Institute of Technology, Atlanta, GA 30332.
*e-mail: jet@ce.gatech.edu

Treating urban populations as "agents of change"[14], with rapidly fluctuating patterns of activities, makes it possible to express higher levels of dynamics than the simple physical locations identified in current master plans[15]. To achieve reliable energy demand predictive models, an approach that incorporates patterns of human activities when quantifying spatial fluctuations of energy use is required. Recent advances in both sensing technologies and urban computing methods have greatly increased the availability of relevant data for urban spaces and supported new discoveries related to these challenges[16-19]. A significant body of work has begun to focus on ways to quantify human activity patterns[17,20,21]. In particular, one of the most popular of the new indicators, human mobility, is now being widely studied. The growing use of humans as sensors has facilitated the collection of city-wide human mobility data[16] via individuals' mobile phone signals, which include GPS data[18,22,23], as has their smart card commuting data[24], and location-embedded information from online social networks[17,25,26], all of which can be used to infer information based on the mobility behavior of urban populations. Here we review statistically significant indications related to the spatial interdependencies between human mobility and urban energy consumption.

**Findings**

We used human mobility as a possible indicator for the induced fluctuations in the spatial distribution of energy use in Greater London, examining human mobility patterns of individuals using 18,810,222 individual positional records from an online social networking platform (Twitter) across 4,835 spatial divisions measuring radius of gyration (*see Methods*). Data from 3,438,939 electricity meters, and 3,007,392 gas meters in the same areas across 33 Greater London boroughs, over the course of 2014 was also used (*Supplementary Table 1-2*). In order to assess the energy use attributable to individuals' urban mobility and thus evaluate the potential utility of human mobility as a predictor of future energy demand, we first examined how human mobility and energy consumption are spatially distributed, including whether there are underlying processes that impose structure on these distributions that can be used to quantify these patterns or are they merely characterized by spatial heterogeneity and randomness (*see Methods*). Figure 1 illustrates these distributions across the 4,835 spatial divisions (referred to in Greater London as Lower Layer Super Output Areas, or LSOAs) for human mobility, electricity and gas consumption.

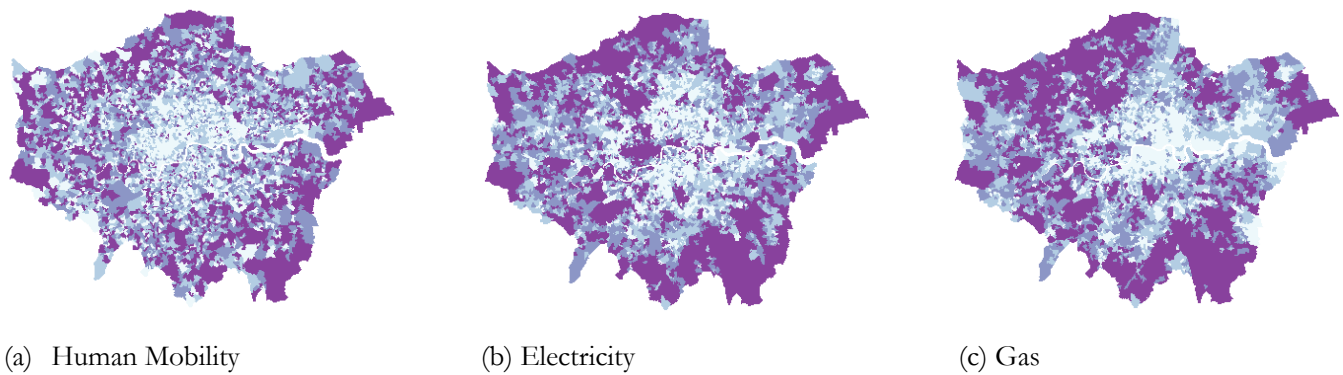

(a) Human Mobility  (b) Electricity  (c) Gas

**Figure 1 | Spatial distribution of human mobility and energy consumption in Greater London, LSOA-level, February 2014. a-c,** human mobility (**a**), electricity consumption (**b**), and gas consumption (**c**).

We found that the spatial distribution of human mobility is not random; an underlying spatial structure governs the mobility of urban population (*Supplementary Results 2.1*). This structure was present throughout the year with only insignificant deviations from the mean (Figure 2). We thus reject the null hypothesis of spatial randomness in favor of structure (i.e., spatial autocorrelation), meaning that the spatial fluctuations of human mobility are relevant and provide additional insights into the structure beyond simply values. Observations of human mobility at one location correlate with those for neighboring locations, with a possible effect on the neighboring values (i.e., values for one division depend on the values at other neighboring locations). Interestingly, this correlation appears to be particularly strong (increased spatial dependency) in September, August, December, and January compared to other months (*Supplementary Table 4, Supplementary Figures 5-6*). The presence of spatial structure suggests that the locations of the individual mobilities' centers of mass (*see Methods: Eq.1*) will be significant, likely as a result of where and how individuals arrange their daily trips to home, work, school, shopping, leisure, and so on. Similar results were obtained for energy (electricity and gas) consumption (*Supplementary Table 3, Supplementary Figure 4*). These spatial autocorrelations suggest predictive models that relate observations of human mobility or energy use at one location to those at other locations can be used to define their particular spatial correlation structure more effectively.

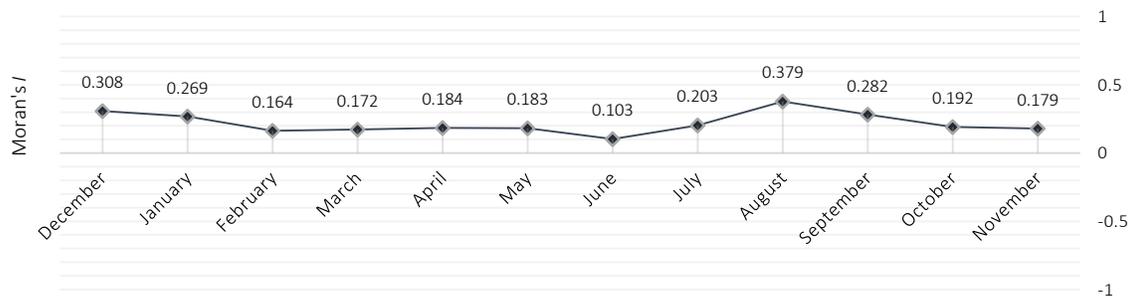

**Figure 2 | Moran's *I*.** Spatial dependence for human mobility by month, 2014.

Once the existence of a spatial structure for both human mobility and energy consumption was confirmed we asked: Is it likely that people's mobility (representing their daily activity patterns) is the cause of the spatial processes (diffusion, interaction, etc.) driving particular energy use patterns in particular locations? If so, does our data support this? Given the spatial autocorrelations, we conducted spatial regression analysis (*See Methods*) to visually (*Supplementary Figures 7-10*) and statistically (*Supplementary Tables 5-12*) explore this hypothesis and determine precisely how the strength of the association between human mobility and energy consumption varies by area. The results of the spatial regression analysis between human mobility and energy use performed to evaluate the contributions of human activities to energy use (electricity and gas) at the urban level revealed that the spatial distribution of energy use was not independent of human mobility; rather the spatial imprints of human mobility localized energy demand distribution. These spatial dependencies were intermittent across the year, reinforcing the finding of an underlying spatial structure for human mobility patterns.

Interestingly, the monthly difference was almost unnoticeable, reinforcing the utility of human mobility as a predictor for urban energy consumption (Figures 3-4). The spatial regression analysis for electricity and gas across Greater London's 4,835 spatial divisions confirmed the existence of statistically significant relationships between human mobility and energy consumption for two spatial autoregressive models (simultaneous autoregressive models consisting of both lag (SAR) and error (SEM) models), with the SAR models predominantly providing the best representations of global dependency conditions (*Supplementary Tables 5-12*). Table 1 depicts the statistical significance and parameters of the predictive SAR models with the lowest Akaike information criterion (*AIC*) for both electricity (*AIC* =77,037) and gas (*AIC* =90,936), which were achieved during the month of February.

**Table 1 | Spatial Autoregressive (SAR) Model.**
Electricity, and gas consumption versus human mobility, February 2014

|  | Spatial Lag Model (SAR) | |
|---|---|---|
|  | Electricity | Gas |
| *AIC for Simple Linear Model (OLS)* | 78920 | 93788 |
| *AIC* | 77037 | 90936 |
| *P-value* | <2.2e-16 *** | <2.2e-16 *** |
| *z-value* | 54.713 | 73.437 |
| *ρ* | 0.69301 | 0.77 |
| *Log likelihood* | -38514.36 | -45463.84 |
| *Approximate Std. Error* | 0.012666 | 0.010485 |

*p<0.05\*; p<0.001\*\*; p<0.0001\*\*\**

The results of the spatial regression analysis indicate that the strength of the association between human mobility and energy consumption depends on spatial location, which can further be contextualized more locally based on Points of Interest (POIs). This means that human mobility across different areas in Greater London can indeed be regarded as a proxy indicator of spatial fluctuations in energy consumption behavior, with changes in human mobility explaining shifts in the pattern of energy consumption that can then be used to quantify and predict spatial flux for energy use of the urban population.

**Discussion and Implications**

Human mobility in urban areas has an undeniable impact on the spatial distribution of energy consumption and can thus serve as a quantitative representation of how an urban population interacts with local energy systems. The results presented in this paper suggest that human mobility can be applied to translate the location-based activities of an urban population into collective energy consumption, thus accounting for the urban energy spatial flux. This study elevates our understanding of the human dimensions of energy use beyond occupants' behavior at the building level[11,27], quantifying a measure of this effect at an urban scale. Human mobility patterns at this wider scale can reveal important information about the way citizens interact with their surroundings, driving energy use. By quantifying these effects, we can measure the strength of relationships, understand interdependencies, and make more reliable predictions of future energy demands.

Our findings regarding the almost invariable spatial dependencies of human mobility over the year as a result of spatial autocorrelation reveal a predictable activity pattern for urban populations and thus energy use within various urban spatial units. Given that the decentralization and efficient allocation of resources is highly dependent on how an urban population's activity patterns are distributed in space, human mobility can be used to infer location choices[28], anticipate future energy demand and strategize optimal decentralization and resource allocation to different amenities under the influence of human mobility, although further research is needed to contextualize this relationship. Knowing how individuals move around urban open spaces and across the physical infrastructure (both communal and private) of an urban landscape will enable us to build a comprehensive understanding of how certain types of energy behavior are clustered in geographical spaces and temporal locations within urban areas. It should be noted that our results do not exclude the possibility that the physical characteristics of the buildings in their spatial context[5-9] also significantly affect urban building energy use. However, developing a comprehensive model of location-based human activity patterns of urban populations by applying the concept of human mobility will enable us to extend studies of urban energy demand beyond the simple physical determinants of energy use. As they continue to grow ever larger, our urban areas will inevitably encounter serious challenges as they strive to meet the energy demands of their expanding populations for which our current knowledge is likely to prove insufficient. To cope with the continuing growth in population and the corresponding increase in urban activities, we need to develop a deeper understanding of the root causes of societally significant phenomena such as energy consumption. The relationship between energy use and human mobility is a key factor for creating effective policies for urban areas. Accurate information on the spatial dependence between fluctuating patterns of human mobility and energy use, the result of an underlying social/behavioral process, can help define a predictable structure for urban energy demand. Spatial dependence is the product of an underlying location specific activity process that leads to clusters of mobility patterns. These patterns can potentially be explained by groupings of particular populations with similar activity patterns or daily routines[29]; diffusion processes[30], where individuals in the same spatial divisions influence, acquire information, and adopt specific energy use patterns; spatial interactions[30], where individuals tend to interact with those who are spatially closer to them; dispersal processes, where individuals travel short distances (e.g., home to work) and transfer their knowledge and energy use patterns with them; externalities and spatial spillover effects[31]; and/or socioeconomic factors in spatial units that drive similar behavior.

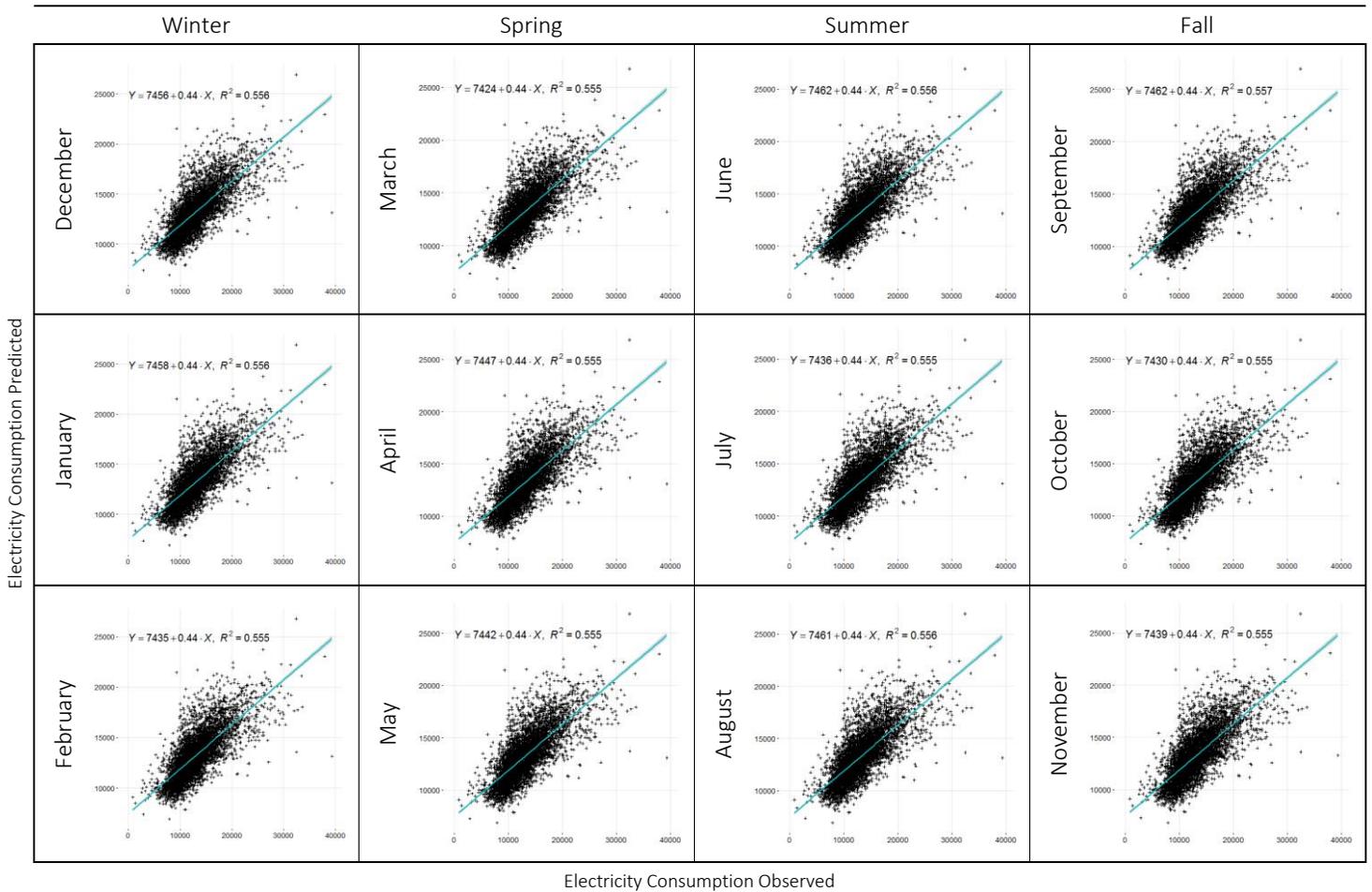

**Figure 3 | Spatial Regression.** Electricity Consumption.

These patterns will also enable us to identify the interdependencies between energy consumption, individual activities, and specific urban spatiotemporal features. Incorporating the spatial imprints into models will advance our understanding and knowledge of the underlying processes and how they propagate across space, shedding new light on the interconnected challenge of theory and analysis.

Perhaps the most striking example of the power of human mobility's impact on urban energy use, and a significant implication of these interdependencies, is the possible spatial spillover effect[31] that determines whether fluctuations in energy use due to human mobility in one spatial unit (i.e., an individual LSOA) have any diffusive impact on its neighboring locations, and if so, whether there is a significant difference in the diffusive effects of these populations. The SAR (simultaneously autoregressive) models, which were found to be the most representative predictive models in this study, permit the magnitude and significance of direct spillover effects to be assessed, showing how changes in human mobility at a particular location will be transmitted to all other locations and thus how they will affect the energy consumption at the corresponding locations.

The availability of such information will allow city managers and policy makers to identify hotspots and develop effective strategies to create bigger energy efficiency spillover effects, or to restrict unwanted or excessive energy use spillover effects. When creating such strategies, individual energy consumption hotspots can be targeted based on the spatial attributes of those locations. Alternatively, particular human mobility networks can become the focus of attention. Diffusing desired effects by introducing changes in the spatial structure (for example by targeting specific buildings or areas to create bigger spillover effects), or instigating contagion by introducing changes in the flow based on contagion (changing the flow, or mobility, by targeting specific clusters of population), will bring urban planners a step closer to achieving better management and allocation of scarce energy resources. The results of this research will also be of value to business practitioners, policy-makers, and research communities by enhancing their future efforts and eliminating overlooked or poorly specified components of urban energy resilience. In particular, by creating a clear picture of the demand-side concentration and diversity, this research will facilitate the appropriate decentralization of the urban energy distribution infrastructure to reduce the vulnerabilities that lead to service disruptions. The main goal of this study has been to contribute to our emerging understanding of how energy use is changing, especially in urban environments. Our ongoing research seeks to understand urban activity patterns across different functional locations using human mobility data to develop integrated predictive models that incorporate temporal elements of activity patterns (for example, recreation, nightlife, shopping, or education) and the resulting fluctuations in the patterns of energy use. Identifying spatial regions with similar temporal activities should allow us to more accurately assess their likely energy use flux and thus optimize the distribution of energy provision.

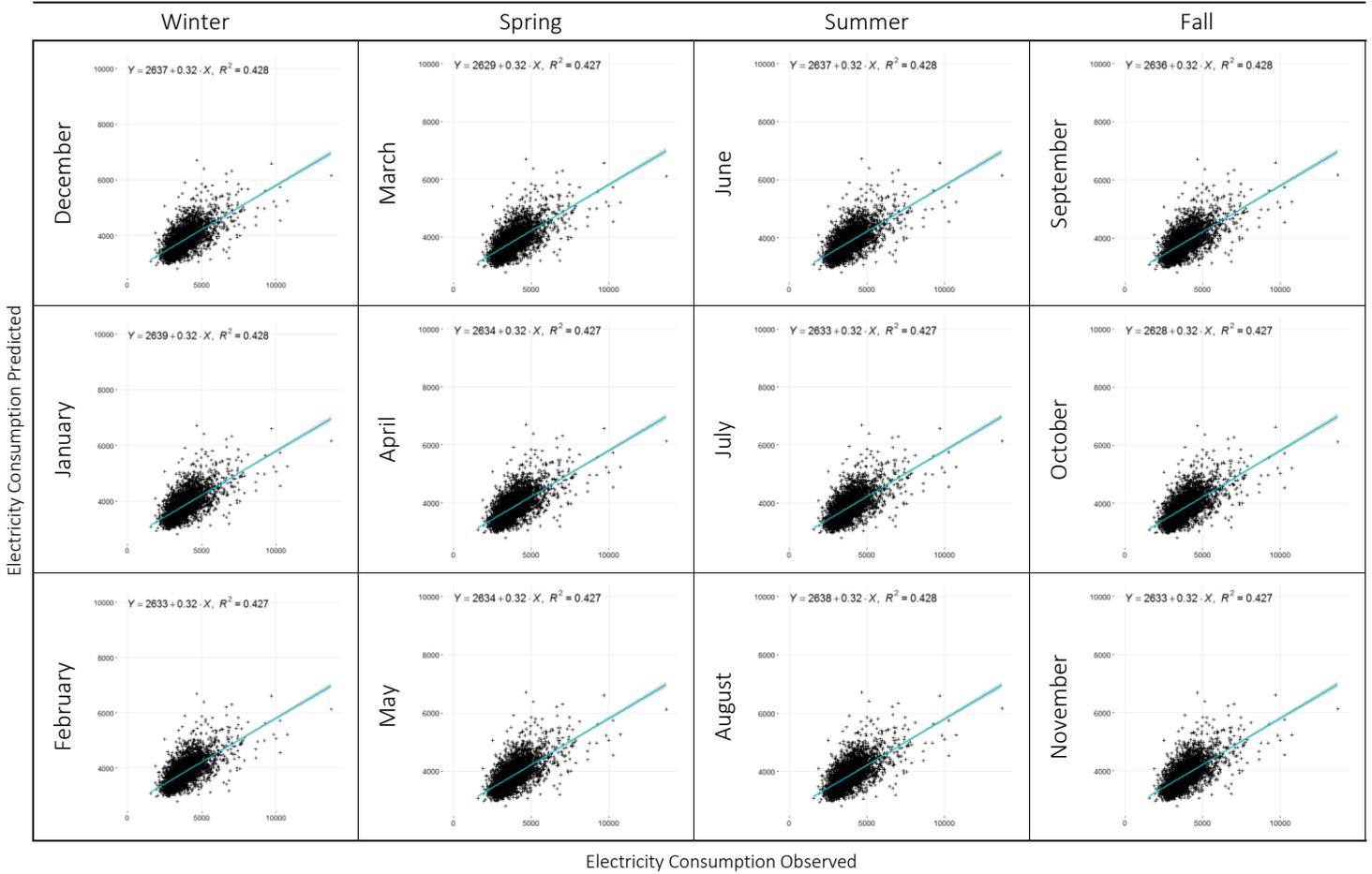

**Figure 4 | Spatial Regression.** Gas Consumption.

This study contributes to efforts to understand how the urban population interacts with local energy systems by linking human mobility patterns to spatial fluctuations of energy use. Knowing individuals' movements around urban open spaces and across the physical infrastructure of our urban environments will enable us to build a comprehensive understanding of how certain types of energy behavior are clustered in specific geographical spaces and temporal locations within urban areas. In addition, it will enable us to identify the interdependencies between energy consumption, individual activities, and specific urban spatiotemporal features. The ability to understand how humans interact with urban energy systems[14] and identify evolving patterns and features in intra-urban mobility routines is important for predicting future patterns of energy demand and protecting energy resilience. Attaining global reductions in energy use and $CO_2$ emissions will demand a paradigm shift in the way we treat energy demand. A clear picture of demand-side diversity that extends beyond the merely physical characteristics of our urban infrastructure will facilitate a more appropriate decentralization of urban energy distribution, thus reducing both, and the vulnerabilities that lead to service disruptions in our ever more complex urban settings.

## Methods
We sought to investigate the interdependencies that may exist between the human activities of an urban population and energy consumption, and, if so, determine whether the distribution of urban energy consumption can be predicted by patterns of human mobility. Using radius of gyration as an indicator for human mobility, we examined the spatial distribution of human mobility and energy use and assessed possible models that could explain the present spatial structure.

**Radius of gyration.** In order to obtain a better understanding of human mobility patterns, the radius of gyration (*Eq. 2*) was selected from among the three most widely accepted indicators used to describe large-scale human mobility patterns: the radius of gyration $r_g(t)$, the trip distance distribution $p(r)$, and the number of visited locations $S(t)$ [22,32,33]. Of these, the radius of gyration was deemed the most appropriate for capturing individuals' characteristic travel distance within the areas where they habitually carry out their daily activities (i.e., $r_{gi}(t)$), as described below:

$$r_{cmi}(t) = \frac{1}{N_{(t)}} \sum_{i=1}^{N_{(t)}} r_i \tag{1}$$

$$r_{gi}(t) = \sqrt{\frac{1}{N_{(t)}} \sum_{i=1}^{N_{(t)}} (r_i - r_{cmi})^2} \tag{2}$$

Here, *N* equals the total number of positional records per individual. The radius of gyration in this study is calculated at two spatial and two temporal levels (*Supplementary Methods 1.2; Supplementary Figure 3*).

**Spatial autocorrelation.** Spatial autocorrelation[34] was used to assess of the extent to which the spatial distribution of the data is compatible with spatial randomness and thus determine whether human mobility and energy consumption do indeed have spatial imprints. Spatial autocorrelation tested the spatial independence of human mobility and energy consumption across 4,835 spatial divisions in Greater London (*Supplementary Results 2.1*). Moran's $I$[35] (*Eq. 5*), which ranges from -1 (most dispersed) to 1 (most clustered), was used to describe the degree of spatial concentration or dispersion for these variables, with large values for $I$ showing clusters of large values that are surrounded by other large values, namely ($I+$)–spatial clustering, and ($I-$)–spatial dispersion, indicating large values that are spatially enclosed by smaller values. It is also a test of independence to determine whether values of human mobility or energy consumption observed in one location depend on the values observed at neighboring locations. While Moran's $I$ represents the global spatial autocorrelation for our data, Geary's $C$[36] (*Eq. 6*) was also used based on the deviations in the responses of each observation with one another, ranging from 0 (maximum positive autocorrelation) to 2 (maximum negative autocorrelation), with 1 indicating an absence of correlation. Moran's $I$ here serves as a measure of sensitivity to extreme values of energy consumption and human mobility, with Geary's $C$ being used to evaluate the sensitivity to differences in smaller neighborhood LSOAs.

$$I = \frac{N\sum_{i=1}^{n}\sum_{j=1}^{n}w_{ij}(x_i - \bar{x})(x_j - \bar{x})}{(\sum_{i=1}^{n}\sum_{j=1}^{n}w_{ij})\sum_{i=1}^{n}(x_i - \bar{x})^2} \quad (5)$$

$$C = \frac{(N-1)\sum_{i=1}^{n}\sum_{j=1}^{n}w_{ij}(x_i - x_j)^2}{2(\sum_{i=1}^{n}\sum_{j=1}^{n}w_{ij})\sum_{i=1}^{n}(x_i - \bar{x})^2} \quad (6)$$

Here, n represents observations on variable x at locations $i, j$ where $\bar{x}$ is the mean of the x variable, and $w_{ij}$ are the elements of the weight matrix. Spatial randomness is undesirable, so to ensure that it is not in effect, we reject the situation of spatial randomness in favor of structure (i.e., spatial autocorrelation). Spatial autocorrelation analysis exactly quantifies this, providing a measure of uncertainty ($p$-value) by which we can reject the null hypothesis (i.e., spatial randomness). A positive spatial autocorrelation indicates that similar values are clusters in neighboring locations, which would be a structure compatible with diffusion[37].

**Spatial regression.** In view of the spatial autocorrelation for human mobility and energy consumption, we investigated the nature of this structure through spatial regression (*Supplementary Results 2.2*). Spatial regression models[38] are used to examine the relationships between variables and their neighboring values and offer a useful way to examine the impact that one observation has on other proximate observations. Starting with an ordinary least square model (*Eq. 7*), with the null hypothesis of a linear regression governing the structure of energy consumption by human mobility as a covariance.

$$y = x\beta + u \quad (7)$$

The expression describes the relationship between a vector of observations on the dependent variable y, a matrix of observations on the explanatory variable x (i.e., human mobility), a vector of regression coefficients $\beta$, and an error term u. The error term is required to have constant variance and must be uncorrelated (i.e., to possess homoscedasticity). While correlations explore the relationships between or among different variables, autocorrelations can be regarded as a special case, as they, explore correlations within variables across space[39]. In the search for an appropriate autocorrelation structure for our data, we tested for deviations that would violate the null hypothesis such as a non-constant variance for error terms (i.e., heteroscedasticity), correlations for the error terms induced by Spatial lag (SAR) (*Eq. 8*), or Spatial Error (SEM) models (*Eq. 9*). The results are shown in Supplementary Tables 5-8 for electricity consumption, and Supplementary Tables 9-12 for gas consumption.

$$y = \rho W y + x\beta + \varepsilon \quad (8)$$

where, $y$ is the dependent variable (i.e., energy consumption: electricity and gas); $x$ is the independent (explanatory) variable (i.e., human mobility); $\beta$ is the regression coefficient; $\varepsilon$ is the random error term; and $\rho$ is the spatial autoregressive coefficient; in the term $\rho W$, which represents the spatially lagged dependent variable.

$$y = X\beta + \lambda W\xi + \varepsilon \quad (9)$$

where, $y$ is the dependent variable (i.e., energy consumption: electricity and gas); $X$ is the independent (explanatory) variable (i.e., human mobility); $\beta$ is the regression coefficient; $\varepsilon$ is the random error term; $\lambda$ is the autoregressive coefficient and $\xi$ represents the normal distribution (0, σ2I) in the term $\lambda W\xi$, which represents the spatial lag for the errors.

**Acknowledgements**

This study was supported by the National Science Foundation under Grant No.1142379. Any opinions, findings, and conclusions or recommendations expressed in this material are those of the authors and do not necessarily reflect the views of the National Science Foundation.